\DocumentMetadata{
      pdfversion=2.0,pdfstandard=ua-2,
    }

\documentclass[sigconf,screen]{acmart}

\usepackage[utf8]{inputenc} 
\usepackage[T1]{fontenc}    
\usepackage{hyperref}       
\usepackage{url}            
\usepackage{booktabs}       
\usepackage{amsfonts}       
\usepackage{nicefrac}       
\usepackage{microtype}      
\usepackage{xcolor}         

\hypersetup{
    colorlinks=true,
    linkcolor=orange,
    filecolor=magenta,      
    urlcolor=orange,
    citecolor=orange,
}

\usepackage{xspace}
\usepackage{mathtools}
\usepackage{bbm}
\usepackage{enumitem}
\usepackage{caption}
\usepackage{subcaption}
\usepackage{multirow}
\usepackage{xcolor}

\usepackage[toc,page,header]{appendix}

\setcitestyle{numbers,square,comma}
\usepackage{amsthm}
\usepackage{amsmath}

\usepackage[capitalise, nameinlink]{cleveref}

\usepackage{xspace}
\usepackage{graphicx}
\usepackage[linesnumbered,ruled,vlined]{algorithm2e}
\usepackage{threeparttable}

\acmDOI{10.1145/3757279.3785638}
\acmYear{2026}
\copyrightyear{2026}
\acmISBN{979-8-4007-2128-1/2026/03}
\acmConference[HRI '26]{Proceedings of the 21st ACM/IEEE International Conference on Human-Robot Interaction}{March 16--19, 2026}{Edinburgh, Scotland, UK}
\acmBooktitle{Proceedings of the 21st ACM/IEEE International Conference on Human-Robot Interaction (HRI '26), March 16--19, 2026, Edinburgh, Scotland, UK}
\received{2025-09-30}
\received[accepted]{2025-12-01}

\begin{document}

\title{DiSCo: Diffusion Sequence Copilots for Shared Autonomy}

\author{Andy Wang}
\authornote{These authors contributed equally to this work.}
\orcid{0009-0007-7210-9936}
\affiliation{%
  \institution{University of California, Los Angeles}
  \city{Los Angeles}
  \country{USA}
}
\email{andywang0321@g.ucla.edu}

\author{Xu Yan}
\authornotemark[1]
\orcid{0000-0003-2493-7693}
\affiliation{%
  \institution{University of California, Los Angeles}
  \city{Los Angeles}
  \country{USA}
}
\email{xkyan3@cs.ucla.edu}

\author{Brandon McMahan}
\authornotemark[1]
\orcid{0009-0009-6798-9670}
\affiliation{%
  \institution{University of California, Los Angeles}
  \city{Los Angeles}
  \country{USA}
}
\email{bmcmahan2025@g.ucla.edu}

\author{Michael Zhou}
\orcid{0009-0008-5826-681X}
\affiliation{%
  \institution{University of California, Los Angeles}
  \city{Los Angeles}
  \country{USA}
}
\email{zhoumichael@g.ucla.edu}

\author{Yuyang Yuan}
\orcid{0009-0008-5911-8659}
\affiliation{%
  \institution{University of California, Los Angeles}
  \city{Los Angeles}
  \country{USA}
}
\email{yuanyuyang@g.ucla.edu}

\author{Johannes Y. Lee}
\orcid{0000-0003-2420-4916}
\affiliation{%
  \institution{University of California, Los Angeles}
  \city{Los Angeles}
  \country{USA}
}
\email{johanneslee@ucla.edu}

\author{Ali Shreif}
\orcid{0009-0009-9892-3455}
\affiliation{%
  \institution{University of California, Los Angeles}
  \city{Los Angeles}
  \country{USA}
}
\email{alimshreif@g.ucla.edu}

\author{Matthew Li}
\orcid{0009-0005-1433-3038}
\affiliation{%
  \institution{University of California, Los Angeles}
  \city{Los Angeles}
  \country{USA}
}
\email{matthewli@g.ucla.edu}

\author{Zhenghao Peng}
\orcid{0000-0001-9726-224X}
\affiliation{%
  \institution{University of California, Los Angeles}
  \city{Los Angeles}
  \country{USA}
}
\email{pzh@cs.ucla.edu}

\author{Bolei Zhou}
\orcid{0000-0003-4030-0684}
\affiliation{%
  \institution{University of California, Los Angeles}
  \city{Los Angeles}
  \country{USA}
}
\email{bolei@cs.ucla.edu}

\author{Yuchen Cui}
\orcid{0000-0001-7417-1222}
\affiliation{%
  \institution{University of California, Los Angeles}
  \city{Los Angeles}
  \country{USA}
}
\email{yuchencui@cs.ucla.edu}

\author{Jonathan C. Kao}
\orcid{0000-0002-9298-0143}
\affiliation{%
  \institution{University of California, Los Angeles}
  \city{Los Angeles}
  \country{USA}
}
\email{kao@seas.ucla.edu}

\renewcommand{\shortauthors}{A. Wang, X. Yan, B. McMahan, M. Zhou, Y. Yuan, J. Y. Lee, A. Shreif, M. Li, Z. Peng, B. Zhou, Y. Cui, and J. C. Kao}

\begin{abstract}
Shared autonomy combines human user and AI copilot actions to control complex systems such as robotic arms.
When a task is challenging, requires high dimensional control, or is subject to corruption, shared autonomy can significantly increase task performance by using a trained copilot to effectively \textit{correct} user actions in a manner \textit{consistent} with the user's goals.
To significantly improve the performance of shared autonomy, we introduce \textbf{Di}ffusion \textbf{S}equence \textbf{Co}pilots (DiSCo): a method of shared autonomy with diffusion policy that plans action sequences consistent with past user actions.
DiSCo seeds and inpaints the diffusion process with user-provided actions with hyperparameters to balance \textit{conformity} to expert actions, \textit{alignment} with user intent, and perceived \textit{responsiveness}.
We demonstrate that DiSCo substantially improves task performance in simulated driving and robotic arm tasks. 
Project website: \url{https://sites.google.com/view/disco-shared-autonomy/}

\end{abstract}



\maketitle

\newcommand{\0}{\mathbf{0}}
\newcommand{\1}{{(1)}}
\newcommand{\2}{{(2)}}
\newcommand{\3}{{(3)}}
\renewcommand{\a}{\mathbf{a}}
\newcommand{\A}{\mathbf{A}}
\renewcommand{\b}{\mathbf{b}}
\newcommand{\bias}{\textrm{bias}}
\newcommand{\B}{\mathbf{B}}
\newcommand{\BN}{\textrm{batch-norm}}
\renewcommand{\c}{\mathbf{c}}
\renewcommand{\c}{\mathbf{c}}
\newcommand{\C}{\mathbf{C}}
\newcommand{\CE}{\textrm{CE}}
\newcommand{\cov}{\mathbf{cov}}
\renewcommand{\d}{\partial}
\newcommand{\D}{\mathcal{D}}
\newcommand{\data}{{\textrm{data}}}
\newcommand{\Dsup}{{(D)}}
\newcommand{\E}{\mathbb{E}}
\newcommand{\e}{\epsilon}
\newcommand{\eps}{\varepsilon}
\newcommand{\f}{\mathbf{f}}
\newcommand{\G}{\mathcal{G}}
\newcommand{\Gsup}{{(G)}}
\newcommand{\g}{\mathbf{g}}
\newcommand{\grad}{\nabla}
\newcommand{\gradlogpi}{\grad_\theta \log \pi_\theta}
\newcommand{\gradlogp}{\grad_\theta \log p_\theta}
\renewcommand{\H}{\mathbf{H}}
\newcommand{\h}{\mathbf{h}}
\newcommand{\hinge}{\textrm{hinge}}
\newcommand{\hmu}{\hat{\mu}}
\newcommand{\htheta}{\hat{\theta}}
\newcommand{\hx}{\hat{\x}}
\newcommand{\hy}{\hat{y}}
\renewcommand{\i}{{(i)}}
\newcommand{\I}{\mathbf{I}}
\newcommand{\ib}{\mathbf{i}}
\newcommand{\intext}{{\textrm{in}}}
\newcommand{\Ind}{\mathbb{I}}
\newcommand\ind{\protect\mathpalette{\protect\independenT}{\perp}}
\newcommand{\J}{\mathbf{J}}
\newcommand{\JS}{\textrm{JS}}
\renewcommand{\j}{{(j)}}
\newcommand{\KL}{\textrm{KL}}
\renewcommand{\L}{\mathcal{L}}
\newcommand{\Lvae}{\L_{\textrm{vae}}}
\newcommand{\leftV}{\left\Vert}
\newcommand{\m}{{(m)}}
\newcommand{\maxout}{\textrm{maxout}}
\newcommand{\model}{{\textrm{model}}}
\newcommand{\MSE}{\textrm{MSE}}
\newcommand{\N}{\mathcal{N}}
\newcommand{\nab}{\nabla}
\newcommand{\new}{{\textrm{new}}}
\renewcommand{\o}{\mathbf{o}}
\renewcommand{\O}{\mathbf{O}}
\newcommand{\out}{{\textrm{out}}}
\newcommand{\opt}{{\textrm{opt}}}
\newcommand{\p}{\mathbf{p}}
\newcommand{\pad}{\texttt{pad}}
\newcommand{\pmodel}{p_{\textrm{model}}}
\newcommand{\pdata}{p_{\textrm{data}}}
\newcommand{\pool}{\texttt{pool}}
\newcommand{\q}{\mathbf{q}}
\newcommand{\rec}{{\textrm{rec}}}
\newcommand{\relu}{\textrm{ReLU}}
\newcommand{\rightV}{\right\Vert}
\newcommand{\range}{\textrm{range}}
\renewcommand{\r}{\mathbf{r}}
\newcommand{\R}{\mathbb{R}}
\newcommand{\calR}{\mathcal{R}}
\newcommand{\s}{\mathbf{s}}
\renewcommand{\S}{\mathbf{S}}
\newcommand{\Si}{\Sigma^{-1}}
\newcommand{\sign}{\textrm{sign}}
\newcommand{\softmax}{\textrm{softmax}}
\newcommand{\softplus}{\textrm{softplus}}
\newcommand{\stride}{\texttt{stride}}
\newcommand{\tausimp}{\tau \sim p_\theta(\tau)}
\newcommand{\SE}{\textrm{SE}}
\newcommand{\tA}{\mathsf{A}}
\newcommand{\thetap}{\theta^\prime}
\newcommand{\tJ}{\tilde{J}}
\newcommand{\tp}{t^\prime}
\newcommand{\tr}{\textrm{tr}}
\newcommand{\tX}{\mathsf{X}}
\newcommand{\tx}{\tilde{\x}}
\renewcommand{\u}{\mathbf{u}}
\newcommand{\U}{\mathbf{U}}
\renewcommand{\v}{\mathbf{v}}
\newcommand{\w}{\mathbf{w}}
\newcommand{\W}{\mathbf{W}}
\newcommand{\var}{\textrm{var}}
\newcommand{\V}{\mathbf{V}}
\newcommand{\X}{\mathbf{X}}
\newcommand{\x}{\mathbf{x}}
\newcommand{\xor}{\textrm{xor}}
\newcommand{\y}{\mathbf{y}}
\newcommand{\Y}{\mathbf{Y}}
\newcommand{\z}{\mathbf{z}}
\newcommand{\Z}{\mathbf{Z}}
\newcommand{\calA}{\mathcal{A}}
\newcommand{\calD}{\mathcal{D}}
\newcommand{\calE}{\mathcal{E}}
\newcommand{\calM}{\mathcal{M}}
\newcommand{\calO}{\mathcal{O}}
\newcommand{\calS}{\mathcal{S}}
\newcommand{\calT}{\mathcal{T}}


\begin{figure*}[!t]
    \centering
    \includegraphics[width=0.9\textwidth]{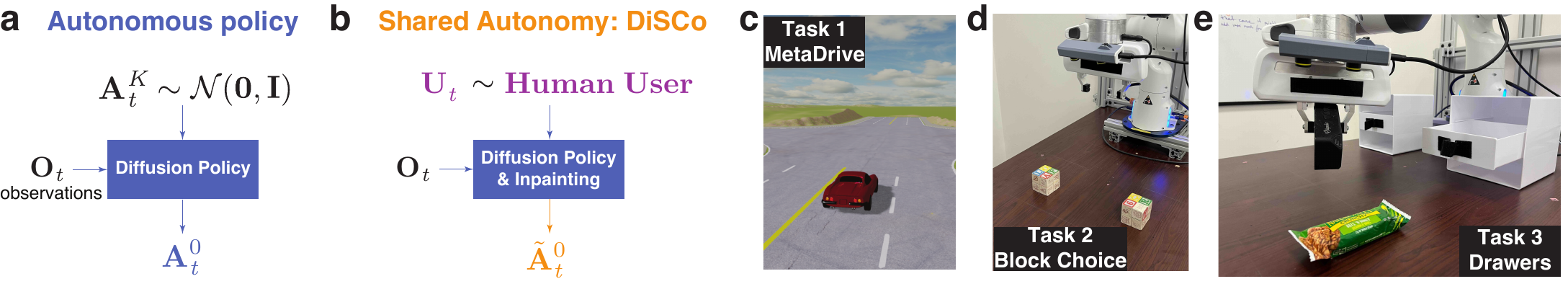}
    \caption{Overview.
    \textbf{a}, Diffusion Policy runs a reverse diffusion process on unit Gaussian noise, conditioned on vision observations $\O_t$, to generate an action sequence $\A_t^0$.
    \textbf{b}, DiSCo diffuses user action sequences with inpainting to generate an action sequence, $\tilde{\A}_t^0$, that reflects user intent.
    \textbf{c, d, e}, We evaluate DiSCo on one simulated task (MetaDrive) and two human-in-the-loop robotic tasks (Block Choice \& Drawers). }
    \label{fig:overview}
    \Description{
        Overview.
        \textbf{a}, Diffusion Policy runs a reverse diffusion process on unit Gaussian noise, conditioned on vision observations $\O_t$, to generate an action sequence $\A_t^0$.
        \textbf{b}, DiSCo diffuses user action sequences with inpainting to generate an action sequence, $\tilde{\A}_t^0$ that reflects user intent.
        \textbf{c, d, e}, We evaluate DiSCo on one simulated task (MetaDrive) and two human-in-the-loop robotic tasks (Block Choice \& Drawers). 
    }
\end{figure*}

\section{Introduction} \label{sec_introduction}

Consider a task where a human user controls a robotic arm to place objects into drawers.
The user has \textit{goals} in mind: she wants to place a particular object -- a Bluetooth speaker -- into a particular drawer.
But in this scenario, she can only control the robotic arm with an underactuated joystick.
This makes it challenging to perform precise high degree-of-freedom actions to execute her goals. 
As a result, her movements are imprecise and suboptimal: they stutter, overshoot, require repeated attempts, and occasionally result in unintended collisions.

While the user has clear \textit{goals} but suboptimal \textit{action execution}, the opposite is true for an AI robot trained to perform the task.
A trained robot's visuomotor policy can be proficient at action execution \cite{chi2023diffusion, Nair2022-so, Ma2022-ra}, but does not know the user's goals.
In prior shared autonomy studies \cite{Javdani2015-db, Gopinath2017-tw, Reddy2018-be, Jun-Jeon2020-nz, Bhattacharjee2020-yb, Losey2022-rd, Tan2022-lf, yoneda2023noise, McMahan2024-sc}, the objective is to combine user actions with trained policies to provide precise action execution toward the user's goals.
This means the robot policy essentially \textit{corrects} or \textit{optimizes} the user's actions while maintaining \textit{alignment} with the user's goals. 
Beyond controlling complex systems, shared autonomy also has applications in brain-computer interfaces (BCIs), where paralyzed participants control robots with neurally decoded signals \cite{Hochberg2012-ys, Collinger2013-aa, Wodlinger2015-ls, Lee2025-ud}.
These decoded control signals are low-dimensional and noisy, effectively corrupting user inputs to the robot.
In these scenarios, shared autonomy may use AI copilots to assist the paralyzed user in executing goals, potentially restoring movement and function \cite{Lee2025-ud}. 

We introduce DiSCo, a diffusion-based shared autonomy framework that (1) enforces user intent via inpainting of action sequences, (2) explicitly balances autonomy, alignment, and responsiveness through interpretable hyperparameters, and (3) achieves increased performance and usability across simulated driving and real-world robotic manipulation tasks under corrupted inputs.
Our method for shared autonomy draws inspiration from diffusion-based image-to-image translation in computer vision. 
In image-to-image translation, a rough sketch undergoes a partial diffusion process, guided by semantic information from text prompts, to yield refined and realistic images \cite{Meng2021-yj, lugmayr2022repaintinpaintingusingdenoising}.
Similarly, DiSCo, which extends the shared autonomy framework by \citet{yoneda2023noise} for Diffusion Policy (Figure~\ref{fig:overview}a) \cite{chi2023diffusion}, transforms noisy or suboptimal user action sequences into optimal action sequences. 
Here, the user's initial action inputs serve as the "rough sketch," while visual context from live multicamera feeds -- encoded by ResNets -- conditions the diffusion process. 
By performing inpainting during reverse diffusion \cite{Lugmayr2022-jl}, DiSCo generates action sequences that harmonize the robot's optimal action execution capabilities with user intentions (Figure~\ref{fig:overview}b).

We evaluate DiSCo and compared its performance against a prior diffusion shared autonomy baseline (\citet{yoneda2023noise}, "State-Based Copilot") and No Copilot on three tasks: (1) MetaDrive, a driving simulator (Figure~\ref{fig:overview}c), (2) Block Choice, where the user must control a robotic arm to pick up one of two blocks (Figure~\ref{fig:overview}d), and (3) Drawers, where the user must place one of two objects into one of two drawers (Figure~\ref{fig:overview}e).
Users inputted commands with a corrupted SpaceMouse, simulating scenarios where user input is far from perfect, and DiSCo therefore has the potential to significantly improve action execution and performance. 
In all tasks, DiSCo significantly increased performance. 
In MetaDrive, DiSCo achieved a success rate of $0.47$ compared to $0.20$ for the State-Based Copilot, with corresponding crash rates of $0.50$ and $0.12$, respectively.
In Block Choice, DiSCo achieved a success rate of $0.80$, compared to $0.45$ with No Copilot and $0.09$ with the State-Based Copilot.
When compared to No Copilot, DiSCo reduced average completion time from $17.8$~s to $14.9$~s and the number of collisions from $4.24$ to $0.15$ per trial.
In Drawers, DiSCo achieved a success rate of $0.42$ versus $0.28$ with No Copilot, reduced average completion time from $93.0$ s to $88.6$ s, and lowered collisions from $9.11$ to $0.74$ per trial. 
Users also perceived DiSCo as easier to control compared to all baselines.

\section{Related Work} \label{sec_related_work}

\noindent \textbf{Diffusion for shared autonomy.}
\citet{yoneda2023noise} used diffusion for shared autonomy, where they trained Denoising Diffusion Probabilistic Models (DDPMs) \cite{Ho2020-lb} conditioned on a state observation $\o_t$ to model a distribution of expert actions, $p(\a_t|\o_t)$. 
They trained this copilot to assist multiple simulation tasks and a robot task of placing a donut-shaped object onto one of two posts.
In their formulation, a user action $\u_t$ undergoes a partial forward diffusion towards unit Gaussian noise by $k_{sw} \leq K$ steps, resulting in $\u_t^{k_{sw}}$.
The partially-noised user action, $\u_t^{k_{sw}}$, then undergoes a partial reverse diffusion to bring it closer to a sample $\a_t \sim p(\a_t|\o_t)$.
The diffusion ratio, $\gamma=k_{sw}/K$, trades off the user's autonomy with conformity to the copilot action distribution.
The critical difference to our work is that this policy only models single time-step actions $p(\a_t | \o_t)$ rather than action sequences, $p(\a_{t-O+1}, \dots, \a_{t+R} | \o_{t-O+1}, \dots, \o_t)$ as in Diffusion Policy (see Methods) \cite{chi2023diffusion}.
For this reason, we call this baseline by \citet{yoneda2023noise} the "State-Based Copilot".
We highlight that (1) generating action sequences involves additional considerations including seeding and inpainting, and (2) DiSCo significantly outperforms this baseline.

\begin{figure*}[t!]
    \centering
    \includegraphics[width=0.95\textwidth]{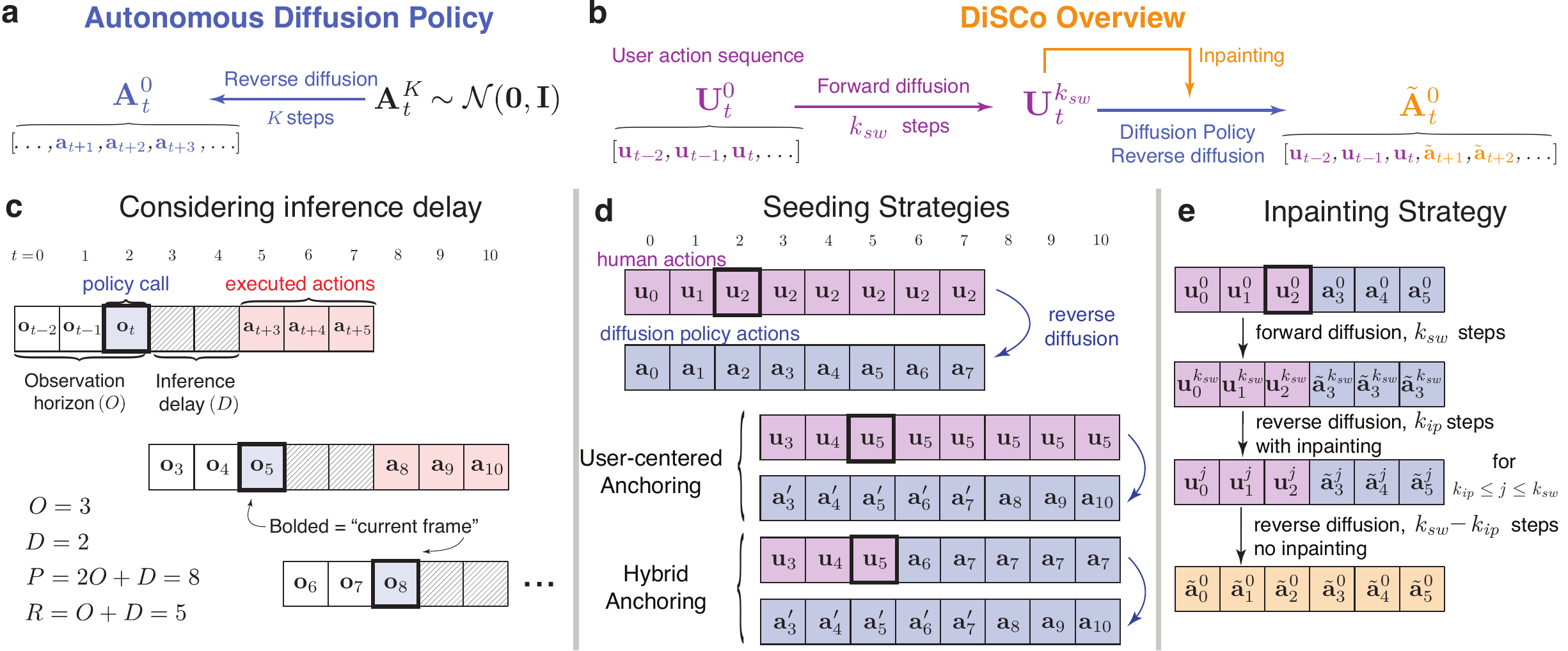}
    \caption{DiSCo overview.
    \textbf{a}, In autonomous diffusion policy, standard Gaussian noise $\A^K_t$ is reverse diffused to an action sequence, $\A_t^0$.
    \textbf{b}, In DiSCo, a user action sequence, $\U_t^0$, is forward diffused by $k_{sw}$ steps to a noisy sequence $\U_t^{k_{sw}}$.
    The noised user inputs $\U_t^{k_{sw}}$ seed the reverse diffusion process and inpaint the reverse process with user actions.
    The output $\tilde{\A}^0_t$ represents the reverse diffused policy action sequence, now inpainted with user actions.
    \textbf{c}, Illustration of the observation horizon $O$, inference delay $D$, re-plan interval $R$, and total prediction horizon $P$. 
    \textbf{d}, The seeding strategies for User-centered Anchoring and Hybrid Anchoring.
    \textbf{e}, The inpainting strategy for shared autonomy.
    After forward diffusion for $k_{sw}$ steps, we run $k_{ip}$ reverse diffusion steps where the the user input sequence is inpainted.
    For the remaining $k_{sw}-k_{ip}$ steps, we run reverse diffusion with no inpainting.
    Total inpainting occurs when $k_{ip} = k_{sw}$, and $\tilde{\a}_0^0 = \u_0^0$, $\tilde{\a}_1^0 = \u_1^0$, etc.
    }
    \Description{DiSCo overview.
    \textbf{a}, In autonomous diffusion policy, standard Gaussian noise $\A^K_t$ is reverse diffused to an action sequence, $\A_t^0$.
    \textbf{b}, In DiSCo, a user action sequence, $\U_t^0$, is forward diffused by $k_{sw}$ steps to a noisy sequence $\U_t^{k_{sw}}$.
    The noised user inputs $\U_t^{k_{sw}}$ seed the reverse diffusion process and inpaint the reverse process with user actions.
    The output $\tilde{\A}^0_t$ represents the reverse diffused policy action sequence, now inpainted with user actions.
    \textbf{c}, Illustration of the observation horizon $O$, inference delay $D$, re-plan interval $R$, and total prediction horizon $P$. 
    \textbf{d}, The seeding strategies for User-centered Anchoring and Hybrid Anchoring.
    \textbf{e}, The inpainting strategy for shared autonomy.
    After forward diffusion for $k_{sw}$ steps, we run $k_{ip}$ reverse diffusion steps where the the user input sequence is inpainted.
    For the remaining $k_{sw}-k_{ip}$ steps, we run reverse diffusion with no inpainting.
    Total inpainting occurs when $k_{ip} = k_{sw}$, and $\tilde{\a}_0^0 = \u_0^0$, $\tilde{\a}_1^0 = \u_1^0$, etc.}
    
    \label{fig:receding}
\end{figure*}

\noindent \textbf{Shared autonomy for robotics.}
In addition to diffusion-based approaches, other works have enabled users to perform goal-oriented tasks via shared autonomy \cite{Javdani2015-db, Gopinath2017-tw, Jun-Jeon2020-nz}.
Most related to this work are shared autonomy studies where the AI copilot helps correct a suboptimal human action \cite{Reddy2018-be, Jun-Jeon2020-nz, Losey2022-rd, yoneda2023noise}.
This includes the use of latent embeddings of high-dimensional robot actions, providing a control space to map low-dimensional inputs to high-dimensional movements \cite{Jun-Jeon2020-nz, Losey2022-rd, cui2023no}, or the use of reinforcement learning to learn high-value actions close to the user's action \cite{Reddy2018-be}.
In contrast, we propose a diffusion process with multiple hyperparameters that enable us to control the degree of \textit{conformity} to the policy action distribution, \textit{alignment} with the user's intent, and real-time \textit{responsiveness} felt by the user.
We note that user autonomy is an important consideration in shared autonomy, as users may prefer control over performance \cite{Javdani2015-db} and autonomous systems  \cite{Bhattacharjee2020-yb}.

\noindent \textbf{Inference-time steering of diffusion policies.}
Recent work adapts pretrained generative controllers without retraining. 
Methods such as ITPS, DynaGuide, and DemoDiffusion \cite{wang2025inferencetimepolicysteeringhuman, du2025dynaguidesteeringdiffusionpolices, park2025demodiffusiononeshothumanimitation} guide diffusion sampling using gradients from task objectives, learned dynamics models, or demonstration priors, respectively. 
These approaches bias generated trajectories toward desired outcomes by softly shaping the reverse diffusion process, which is helpful for goal-conditioned planning and task adaptation. 
These methods primarily operate on autonomous policies and typically assume reliable state observations and low-latency control, making them less suited to settings with corrupted human inputs. 
DiSCo directly integrates user intent into the generative process, rather than indirectly steering samples via auxiliary objectives
and is better suited for shared autonomy, where preserving alignment with user intent and maintaining real-time responsiveness are critical.

\section{Methods} \label{sec_methods}

We first provide background on Diffusion Policy \cite{chi2023diffusion}. 
We then detail DiSCo and our task settings. 

\subsection{Background: Diffusion Policy}

Diffusion policy is a visuomotor policy based on DDPM \cite{Ho2020-lb}.
At time step $t$, diffusion policy is conditioned on a sequence of $O$ observations, $\O_t = \{\o_{t-O+1}, \dots, \o_t\}$ to predict an action sequence $\A_t^0 = \{\a_{t-O+1}, \dots, \a_t, \dots, \a_{t+R}\}$ of length of $P=O+R$.
Instead of modeling $p(\A_t, \O_t)$, diffusion policy uses DDPM to approximate the conditional distribution  $p(\A_t | \O_t)$ to predict actions conditioned on observations.
At iteration $k$, diffusion policy computes:
\begin{eqnarray}
    \A_t^{k-1} = \alpha \left(\A_t^k - \gamma \epsilon_\theta\left(\O_t, \A_t^k, k) + \N(\0, \sigma^2\I\right) \right),
\end{eqnarray}
where $\alpha, \gamma$ are constants, $\epsilon_\theta$ is a trained neural network with parameters that predicts the noise to subtract to infer $\A_t^{k-1}$, and $\N(\0, \sigma^2\I)$ is Gaussian noise added at each iteration. 
The parameters, $\theta$, are optimized to minimize the mean squared error (MSE),
\begin{eqnarray}
    \L = \mathrm{MSE}\left(\epsilon^k, \epsilon_\theta\left(\O_t, \A_t^0 + \epsilon^k, k\right)\right),
\end{eqnarray}
where $\epsilon^k$ is the added noise at iteration $k$.
To sample an action sequence $\A_t^0 \sim p(\A_t | \O_t)$ involves running $K$ reverse diffusion steps from unit Gaussian noise $\A_t^K \sim \N(\0, \I)$ (Figure~\ref{fig:receding}a). 
This incurs an inference delay of duration $D$.
This inference delay can be reduced by using other sampling methods \cite{Song2020-jh}, although at the potential cost of lower quality samples compared to DDPM.

Diffusion policy is trained end-to-end with data pairs $\{\O_t, \A_t\}$.
\citet{chi2023diffusion} presented two architectures, one based on a CNN that conditions on $\O_t$ using Feature-wise Linear Modulation (FiLM) and a transformer-based architecture.
We adopt a CNN-based diffusion policy rather than a Transformer to reduce real-time inference latency, which is critical for responsive shared autonomy.

\subsection{Considerations for Shared Autonomy with Diffusion Policy}

\textbf{Receding horizon control.} 
In real-time control, diffusion policy may exhibit observable pauses because it incurs a delay to sample $\A_t^0 \sim p(\A_t | \O_t)$ through reverse diffusion. 
We refer to these execution delays as "stuttering."
Stuttering should be minimized or eliminated in shared autonomy, where it is critical to preserve user autonomy and reduce frustration \cite{Javdani2015-db, Bhattacharjee2020-yb, McMahan2024-sc}.
Consider a diffusion policy with observation horizon, $O$, and an upper bound for inference delay, $D$ (Figure~\ref{fig:receding}a).
Let $R$, the "replan interval," be the number of actions planned after time $t$, i.e., $\{\a_{t+1}, \dots, \a_{t+R}\}$.
When $R < O+D$, stuttering occurs because the reverse process may not complete before the next inference step. 
To avoid this, we require $R \geq O+D$.
We set $R = O+D$ to minimize the system loop time. 
Under these assumptions, the prediction horizon for diffusion policy is $P=O+R=2O+D$.
This is illustrated in Figure~\ref{fig:receding}a.
We note that the reverse diffusion process can be run every $O$ frames for computational efficiency.
In our implementation, reverse diffusion incurs an average inference delay of 356 ms, motivating architectural and blending choices described below.
In our experiments, we used: $O=6$, $D=4$, so that $P=16$, where each frame was $100$~ms.
The total number of diffusion steps was $K=100$. 

\textbf{Seeding strategy.} 
In autonomous diffusion policy, the reverse process is run on $\A_t^K=\left[\x_{t-O+1}, \dots, \x_t, \dots, \x_{t+R}\right]$ where each $\x_\tau~\sim~\N(\0,\I)$ for $\tau=t-O+1, \dots, t+R$.
In shared autonomy, when diffusion policy is called at time $t$, the user has provided input actions up to and including time $t$, $\{\u_{t-O+1}, \dots, \u_{t}\}$. 
The user's actions may therefore seed the diffusion process.
Adapting the approach of \citet{yoneda2023noise}, we seed a pilot action sequence, $\U_t^0$ and perform $k_{sw}$ forward diffusion steps to compute a noised action sequence, $\U_t^{k_{sw}}$ (Figure~\ref{fig:receding}b).

Because we are generating a $P$-length action sequence, we must seed the reverse diffusion process with future actions.
We propose two seeding strategies for $\U_t^0$, illustrated in Figure~\ref{fig:receding}d for $O=3$ and $D=2$.
In "User-centric Anchoring," the reverse diffusion process is seeded solely with user actions. 
Because we do not have observations of the user's future actions, we repeat the action $\u_t$ for time steps $\tau=t+1, \dots, t+R$, so that for User-centric Anchoring, $\U_t^0 = \{ \u_{t-O+1}, \dots, \u_t, \dots, \u_t \}$.
In "Hybrid Anchoring," we seed using the last action sequence output, $\A_{t-1}^0$. 
Because the last action sequence was inferred at time $\tau = t-O+1$, this means the last inferred action is $\a_{t+R-O}$.
For Hybrid Anchoring, we therefore have that $\U_t^0 = \{\u_{t-O+1}, \dots, \u_t, \a_{t+1}, \dots, \a_{t+R-O}, \dots, \a_{t+R-O}\}$, where the last inferred action $\a_{t+R-O}$ is repeated until time $t+R$ (Figure~\ref{fig:receding}d).

\textbf{Inpainting user actions into diffusion policy.}
In reverse diffusion, inpainting harmonizes distribution samples with user inputs \cite{Lugmayr2022-jl}.
We inpaint the reverse diffusion process using $\U_t^{k_{sw}}$ (Figure~\ref{fig:receding}b).
We define an "inpainting ratio," $\rho \in [0,1]$ quantifying the proportion of reverse diffusion steps with inpainting. 
Then, $\rho = k_{ip} / k_{sw}$, where $k_{ip}$ is the number of reverse diffusion steps with inpainting, illustrated in Figure~\ref{fig:receding}e.
Note that $k_{ip} = \rho k_{sw} = \rho \gamma K$.
User actions are inpainted for $k_{ip}$ steps until $\A_t^{k_{sw}-k_{ip}}$.
Then for the remaining reverse diffusion steps towards $\A_t^0$, there is no inpainting.
When $\rho=1$, there is total inpainting, and when $0<\rho<1$, there is partial inpainting.

We draw an analogy to computer vision for intuition on $\gamma$ and $\rho$. 
$\gamma$, which controls how many timesteps to forward diffuse, is analogous to SDEdit \cite{Meng2021-yj} while $\rho$, which controls how many reverse diffusion timesteps are inpainted, is analogous to RePaint \cite{Lugmayr2022-jl}.
The diffusion and inpainting ratios $\gamma$ and $\rho$ can therefore be interpreted as controlling the \textit{conformity} and \textit{alignment} of DiSCo. 
When $\gamma$ is larger, the action sequence is closer to a sample from diffusion policy. 
Large values of $\gamma$ therefore help to correct errant user input, although the policy's corrective actions may be towards a goal the user did not intend.
To counteract this, $\rho$ can be used to ensure greater alignment with the user actions.
Larger $\rho$ increases the amount of inpainting in the reverse diffusion process, causing the inferred actions to be more consistent with the user input.
For example, we show it is possible to generate actions consistent with the user's intended goals even when $\gamma=1$, so long as $\rho$ is also close to $1$, since inpainting harmonizes actions to be consistent with the user's actions.

\textbf{Blending user and DiSCo actions.}
Users may prefer control over performance \cite{Javdani2015-db}.
We therefore include a simple "blending ratio", $\beta \in [0,1]$, that blends the user and DiSCo action via a convex combination. 
Let $\u_t$ and $\tilde{\a}_t$ be the user and DiSCo action at time $t$, respectively.
We compute $\tilde{\a}_t \leftarrow \beta \u_t + (1-\beta) \tilde{\a}_t$.
When $\beta=1$, this corresponds to user control with no copilot.
Setting $\beta>0$ improves responsiveness and increases user control, since the user's current action is played immediately, as opposed to via reverse diffused actions after $D$ frames.
In MetaDrive, responsiveness is not an issue because the environment is simulated, and we set $\beta=0$.
However, in real-time robotic arm tasks, we found users preferred a non-zero $\beta$.
Our experiments use $\beta=0.3$ in Block Choice and Drawers. 


\subsection{Tasks and Data Collection}

We evaluated DiSCo in three tasks, each comparing to pilot only control ("No Copilot", $\beta=1$) and the \citet{yoneda2023noise} baseline ("State-Based Copilot").
All policies were evaluated in a random interleaved order.
The user study involved in this paper was approved by the UCLA Institutional Review Board.
All visual observations for the robotic arm tasks were captured using ZED stereo cameras. 
All experiments used a Franka Emika Panda robot.

\textbf{Corruption of user actions.}
To evaluate DiSCo's ability to improve control in challenging settings, we corrupted user actions.
The corruptions were inspired by the BCI literature, where disabled individuals control robotic arms through neurally decoded signals \cite{Hochberg2012-ys, Collinger2013-aa, Wodlinger2015-ls}.
When neural signals are non-invasive \cite{Lawhern2018-fi, Forenzo2024-ue}, they have significant temporal lag and are frequently erroneous \cite{Lee2025-ud}.
We therefore corrupted the user action as follows: (1) with $p=0.95$, we delayed $\u_t$ by three frames ($300$~ms), and (2) with $p=0.5$, we negated an element of $\u_t$, reflecting noisy action in the incorrect direction.
For MetaDrive, the user actions were $\u_t \in \R^2$ quantifying the steering wheel position and throttle, each clipped from $[-1, 1]$. 
For robotic arm control, the user actions were $\u_t \in \R^7$ corresponding to $\{\Delta x, \Delta y, \Delta z\}$ position, $\{\Delta \textrm{roll}, \Delta \textrm{pitch}, \Delta \textrm{yaw}\}$, and $\Delta \textrm{gripper}$ state.

\textbf{Task 1: 4-Goal MetaDrive (simulated).} 
We built a MetaDrive environment \cite{Li2023-xm} with a 4-way intersection (Figure~\ref{fig:overview}c), called "4-Goal MetaDrive".
To complete the task, the car had to drive to the intersection and successfully execute one of the four potential goals: (1) making a U-turn, (2) making a left turn, (3) making a right turn, or (4) going straight.
The car had to then drive to the end of the road.
We trained an expert $Q$-agent to simulate a corrupted surrogate human policy.
Successful trials corresponded to episodes where the car executed the correct goal; if the car arrived at an incorrect goal, it was not successful.
Crash trials corresponded to episodes where the vehicle went off-road.
Because MetaDrive uses a surrogate policy and simulated environment, we were not impacted by inference delay and therefore did not evaluate seeding strategies or blending.

\begin{figure*}
    \centering
    \includegraphics[width=0.9\textwidth]{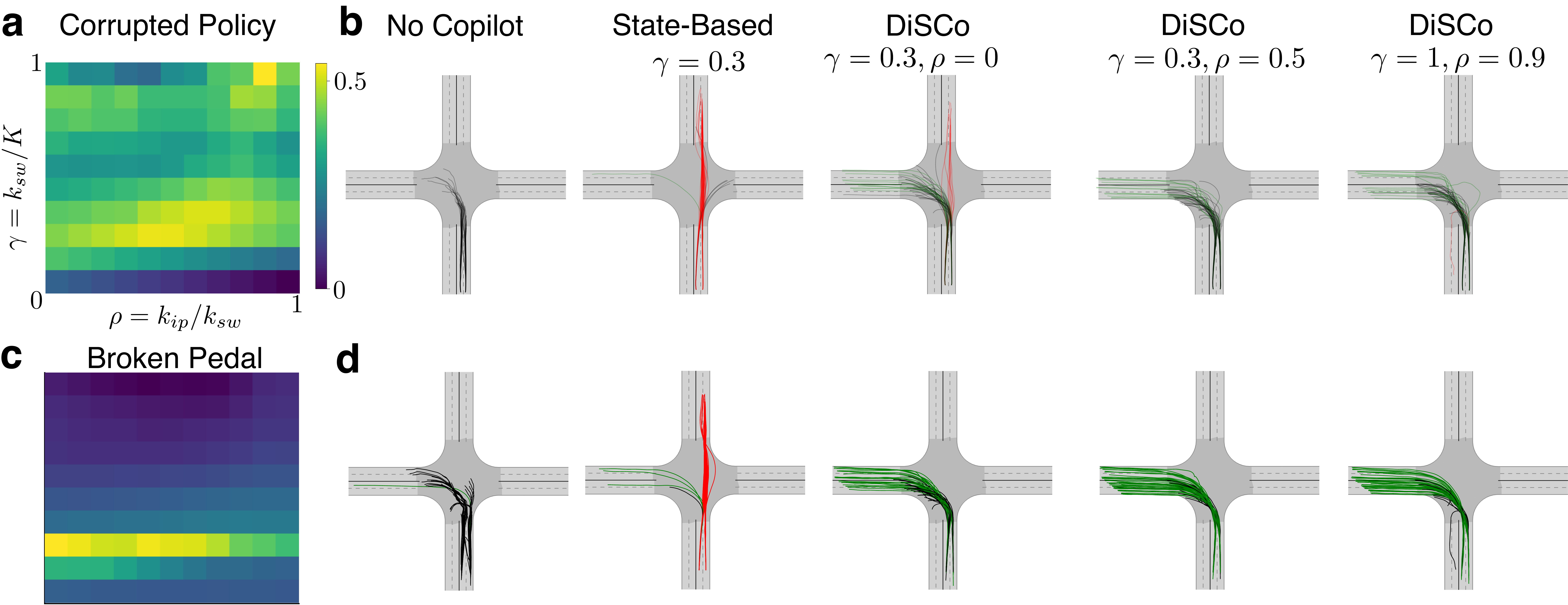}
    \caption{4-Goal MetaDrive hyperparameter sweep and trajectories.
    \textbf{a}, Hyperparameter sweep of $\gamma$ and $\rho$ for 4-Goal MetaDrive using the Corrupted Policy.
    \textbf{b}, Example trajectories where the correct goal was to turn left.
    The No Copilot condition repeatedly crashed (black).
    The State-Based Copilot condition had non-crash trajectories that reached the wrong goal (red).
    DiSCo, at varying levels of $\gamma$ and $\rho$, were able to correctly achieve the goal (green).
    \textbf{c, d}, Same as (a, b) but for the "Broken Pedal" policy where the pedal is zeroed with probability $0.85$.
    }
    \label{fig:metadrive}
    \Description{4-Goal MetaDrive hyperparameter sweep and trajectories.
    \textbf{a}, Hyperparameter sweep of $\gamma$ and $\rho$ for 4-Goal MetaDrive using the Corrupted Policy.
    \textbf{b}, Example trajectories where the correct goal was to turn left.
    The No Copilot condition repeatedly crashed (black).
    The State-Based Copilot condition had non-crash trajectories that reached the wrong goal (red).
    DiSCo, at varying levels of $\gamma$ and $\rho$, were able to correctly achieve the goal (green).
    \textbf{c, d}, Same as (a, b) but for the "Broken Pedal" policy where the pedal is zeroed with probability $0.85$.}
\end{figure*}

\textbf{Task 2: Robotic Arm Block Choice.}
In "Block Choice," participants were instructed to use a SpaceMouse (3Dconnexion) to control the robotic arm to pick up one of two blocks on a table within $20$~seconds (Figure~\ref{fig:overview}d). 
Participants were given 5 minutes to practice controlling the robot before trials began.
Correct (wrong) trials corresponded to trials where the user picked up the correct (incorrect) block.
"Collisions" quantify the average number of times per trial the robotic arm collided with the table or block.
We conducted a user study with 10 participants.
We also quantified each user's perceived "Ease," "Control," and "Effect" on a scale of 1 (worst) to 5 (best) for each condition (see Appendix A for additional details). 
To provide finer granularity and better align with established measures in HRI research, we additionally administered five NASA-TLX–based \cite{hart1988development} items reflecting the perceived workloads during the second phase of the study (see Appendix A for details). 
The autonomous diffusion policy was trained after collecting 200 demonstrations using a SpaceMouse.
The block locations were randomized for each trial.

\textbf{Task 3: Robotic Arm Drawers Task.}
In "Drawers," participants were instructed to use the SpaceMouse to control the robotic arm to move one of two objects (a granola bar or a Bluetooth speaker) into one of two drawers within $100$~seconds (Figure~\ref{fig:overview}e). 
A trial was successful if the user picked up the correctly prompted object and placed it in the correctly prompted drawer.
This involved a sequence of navigating the robotic arm to the correct drawer, opening it, then navigating the robotic arm to the correct object, picking it up, and then navigating the held object to the correct drawer, subsequently placing it correctly in the drawer.
This task represents a more challenging, temporally extended shared-autonomy setting, requiring sustained intent inference and assistance across multiple sequential subgoals, analogous to multi-stage BCI manipulation tasks.
Collisions quantify the average number of times the robotic arm collided with the table, object, or drawer per trial.
We conducted a user study with 10 participants, and also quantified each user's subjective perception using the same set of questions as Task 2. 
The autonomous policy was trained after collecting 120 demonstrations using a SpaceMouse.
We collected an equal number of trials of all $4$ possible object-drawer permutations.
The robotic arm starting position was random in the training demonstrations.

Together, these tasks reflect increasing demands on shared autonomy, from continuous target-directed control to temporally extended, multi-goal manipulation.

\section{Experiments}
\label{sec:result}

Unless otherwise stated, we use Kruskal–Wallis tests for multi-condition comparisons and Wilcoxon rank-sum tests for pairwise comparisons, with significance assessed at $\alpha=0.01$.
In robotic arm statistics, we first applied a Kruskal–Wallis test across all policy conditions to assess overall differences; when significant, we performed pairwise comparisons using the Wilcoxon rank-sum test for both objective performance metrics and subjective ratings, treating trials as independent samples.

\begin{figure*}
    \centering
    \includegraphics[width=0.95\textwidth]{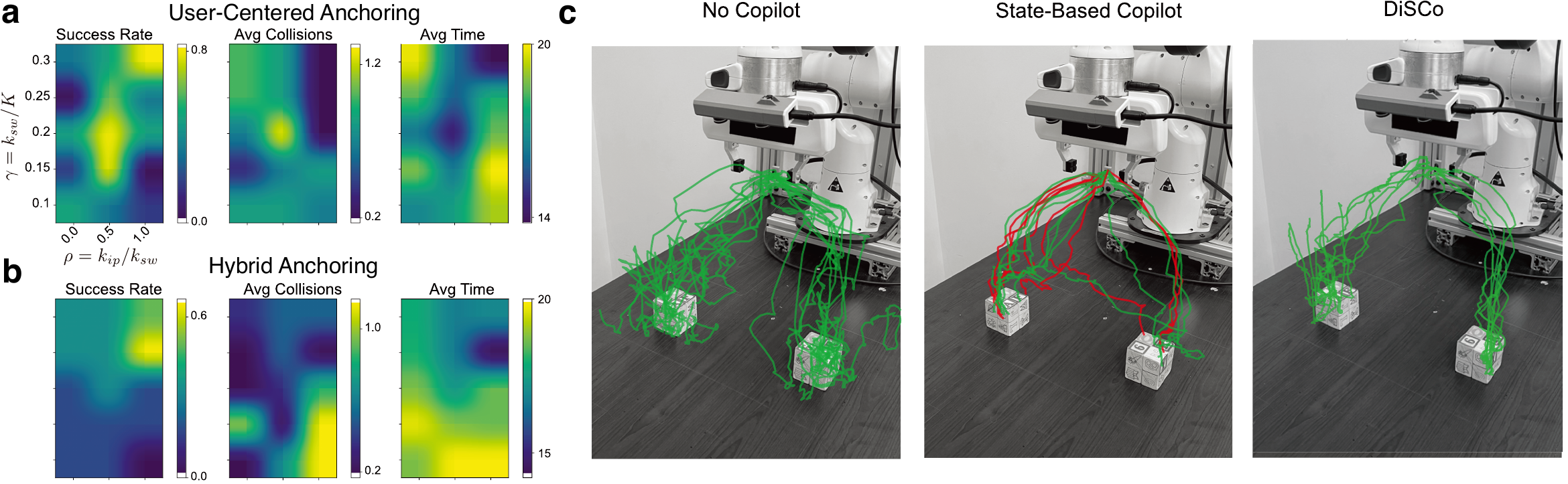}
    \caption{Block Choice hyperparameter sweep and trajectories.
    \textbf{a, b}, Hyperparameter sweeps for User-Centered Anchoring and Hybrid Anchoring seeding strategies.
    \textbf{c}, Sample trajectories of block choice for User 2 under No Copilot, State-Based Copilot ($\gamma = 0.3$), and DiSCo ($\gamma = 0.3$, $\rho=0.1$, $\beta=0.3$) control.
    Green lines correspond to trajectories of correct trials, while red lines correspond to trajectories of wrong trials. 
    }
    \label{fig:blocks}
    \Description{Block Choice hyperparameter sweep and trajectories.
    \textbf{a, b}, Hyperparameter sweeps for User-Centered Anchoring and Hybrid Anchoring seeding strategies.
    \textbf{c}, Sample trajectories of block choice for User 2 under No Copilot, State-Based Copilot ($\gamma = 0.3$), and DiSCo ($\gamma = 0.3$, $\rho=0.1$, $\beta=0.3$) control.
    Green lines correspond to trajectories of correct trials while red lines correspond to these of wrong trials.}
\end{figure*}

\subsection{Simulation Environment: 4-Goal MetaDrive}

\begin{table}[b!]
\vspace{-1em}
\centering
\caption{Comparison of performance metrics under different copilot configurations. Each condition was evaluated over 50 episodes 10 times and we found all DiSCo conditions had significantly higher success rate than others ($p < 0.01$, Wilcoxon rank-sum test). }
\begin{threeparttable}
\begin{tabular}{c|ccc}
\toprule
Policy & 
Success Rate & 
Crash Rate \cr 
\midrule

No Copilot                          & 0.002        & 0.998        \cr
State-Based Copilot ($\gamma=0.3$)        & 0.200       & \textbf{0.120}     \cr
DiSCo ($\gamma=0.3$, $\rho=0.0$)    & 0.381    & 0.386     \cr
DiSCo ($\gamma=0.3$, $\rho=0.5$)    & \textbf{0.471}    & 0.500        \cr
DiSCo ($\gamma=1.0$, $\rho=0.9$)    & 0.398   & 0.596    \cr
\bottomrule
\end{tabular}
\end{threeparttable}

\label{tab:metadrive}
\end{table}

How do $\gamma$ and $\rho$ affect performance of DiSCo?
We performed a grid search in 4-Goal MetaDrive and quantified the fraction of successfully completed goals (Figure~\ref{fig:metadrive}a). 
We found two optimal regions.
One was a relatively wide optimum with $\gamma$ between $0.2$ to $0.3$, and $\rho$ between $0.3$ to $0.9$.
The other was at $\gamma\approx1$ and $\rho\approx0.9$.
The optimal around $\gamma\approx0.2$ is consistent with findings by \citet{yoneda2023noise} for a State-Based Copilot.
But surprisingly, there was an optimum at $\gamma=1$, where future user actions are diffused from unit Gaussian noise.
In the State-Based Copilot, this copilot would contain no information about user goals.
However, because DiSCo inpaints past user actions at every reverse diffusion step with $\rho=0.9$, predicted actions for past timesteps are \textit{consistent} with past user actions.
It is therefore possible for DiSCo to provide higher \textit{correction} (larger $\gamma$) while still enabling \textit{alignment} (larger $\rho$) with user actions.
This is illustrated in sample trajectories in Figure~\ref{fig:metadrive}b. 
In another surrogate policy where we simulated a broken pedal, we found similar results, although DiSCo with $\gamma\approx0.3, \rho\approx0.5$ performed significantly better.

We evaluated success and crash rate in Table~\ref{tab:metadrive}.
We compare conditions using the Wilcoxon rank-sum test, as trials across policies are independent.
We found that DiSCo outperformed No Copilot and the State-Based Copilot in goal success rate ($p<0.01$, Wilcoxon ranksum test) while also avoiding crashes compared to No Copilot ($p<0.01$, Wilcoxon ranksum test).
Note, the crash rate of the State-Based Copilot is low, in part because it frequently drove straight at the intersection even when it was not the correct goal (Figure~\ref{fig:metadrive}b,d), avoiding turns.
Together, these results demonstrate that DiSCo outperforms No Copilot and State-Based Copilot conditions in a driving simulator. 


\subsection{Robotic Arm Block Choice}

We performed an empirical hyperparameter sweep of $\gamma$ and $\rho$ for both User-centered Anchoring (Figure~\ref{fig:blocks}a) and Hybrid Anchoring (Figure~\ref{fig:blocks}b), evaluating 6 trials per condition.
User-centered Anchoring achieved higher success rates and lower average trial times than Hybrid Anchoring.
Anecdotally, we observed that Hybrid Anchoring was less responsive to user actions, and could result in compounding error if the diffusion policy predicted suboptimal actions that then seeded the reverse diffusion process.
In all Block Choice experiments, we used User-centered Anchoring, with $\gamma=0.3$, $\rho=1$, and $\beta=0.3$.
This provided modest conformity to user actions ($\gamma=0.3$) while constraining policy actions to harmonize with total inpainted user actions ($\rho=1$) and providing responsiveness ($\beta=0.3$).
As our data distribution does not meet ANOVA assumptions, we conducted Kruskal-Wallis tests and found statistically significant differences in all metrics ($p < 0.01$, Table~\ref{tab:blocks-stats}).
As shown in Table~\ref{tab:blocks}, DiSCo achieved higher success rates, fewer wrong trials, and significantly faster acquisition times compared to the State-Based Copilot ($p < 0.01$, Wilcoxon ranksum test, Table~\ref{tab:blocks}).
DiSCo also significantly reduced collisions relative to the No Copilot condition ($p < 0.01$, Wilcoxon ranksum test).

\begin{table}[b!]
\vspace{-1em}
\centering
\caption{Kruskal–Wallis test results for \textbf{Block Choice} dataset ($\alpha = 0.01$).
Statistics across $n=10$ participants.}
\label{tab:blocks-stats}
\begin{tabular}{lccc}
\toprule
Measure & $H$ & $p$ & $p < 0.01$ \\
\midrule
Success        & 12.764  & $5.2\times 10^{-3}$ & Yes \\
Time (s)       & 133.321 & $1.0\times 10^{-28}$ & Yes \\
Collisions     & 346.107 & $1.0\times 10^{-74}$ & Yes \\
Wrong          & 55.216  & $6.2\times 10^{-12}$ & Yes \\
Ease           & 139.892 & $4.0\times 10^{-30}$ & Yes \\
Control        & 171.022 & $7.7\times 10^{-37}$ & Yes \\
Effect         & 176.105 & $6.1\times 10^{-38}$ & Yes \\
Mental Demand  & 28.824  & $5.5\times 10^{-7}$  & Yes \\
Physical Demand& 37.357  & $7.7\times 10^{-9}$  & Yes \\
Effort         & 33.733  & $4.7\times 10^{-8}$  & Yes \\
Performance    & 81.719  & $1.8\times 10^{-18}$ & Yes \\
Frustration    & 29.003  & $5.0\times 10^{-7}$  & Yes \\
\bottomrule
\end{tabular}
\end{table}

\begin{table*}[t]
\vspace{-2em}
\caption{Comparison of performance on Block Choice across all policy conditions.}
\begin{threeparttable}
\begin{tabular}{c|ccccccc}
\toprule
Policy & Success $\uparrow$ & Wrong $\downarrow$  & Time (s) $\downarrow$ & Collisions $\downarrow$ & Ease $\uparrow$ & Control $\uparrow$ & Autonomy $\uparrow$ \cr
\midrule
No Copilot & 0.45 & \textbf{0.00} & 17.82 & 4.24 & 2.80 & 3.14 & 3.26 \cr
State-Based Copilot & 0.09 & 0.23 & 19.97 & 0.25 & 1.72 & 1.90 & 1.94 \cr
DiSCo & \textbf{0.80} & \textbf{0.00} & \textbf{14.90} & \textbf{0.15} & \textbf{3.62} & \textbf{3.78} & \textbf{4.04}  \cr
\bottomrule
\end{tabular}
\end{threeparttable}
\label{tab:blocks}
\end{table*}

While DiSCo empirically improved all metrics, in what ways did DiSCo improve performance on these tasks?
To assess this, we plot representative sample trajectories from one user, shown in Figure~\ref{fig:blocks}c.
First, we observed that copilots helped to significantly smooth trajectories, consistent with policy actions conforming more closely to a sample from diffusion policy.
Both the State-Based Copilot and DiSCo had smoother trajectories than the No Copilot condition. 
Along with these generally smoother trajectories, we correspondingly observed that both the State-Based Copilot and DiSCo significantly reduced collisions with the block or table compared to No Copilot (Table~\ref{tab:blocks}, "Collisions").
Second, we observed that DiSCo had significantly higher success rates to the correct goal than the State-Based Copilot (green trajectories in Figure~\ref{fig:blocks}c).
The State-Based Copilot, which inferred only a single action given a single observation, occasionally produced trajectories towards the wrong block (red trajectories in Figure~\ref{fig:blocks}c), even as the user would supply inputs that were consistent with the correct goal and inconsistent with the State-Based Copilot incorrect movements. 
As a result, the State-Based Copilot chose the wrong target $23\%$ of the time (compared to $0\%$ for DiSCo, Table~\ref{tab:blocks}, "Wrong").
Users therefore reported that they felt less control and ease when using the State-Based Copilot (Table~\ref{tab:blocks}, "Ease" and "Control").
In contrast, DiSCo never incorrectly inferred the goal of the user, as it never reached to the incorrect block.
DiSCo therefore more accurately inferred the correct goal of the user.
Third, we observed that the State-Based Copilot only had successful block reaches on $9\%$ of the trials, while DiSCo was successful on $80\%$ of the trials.
We note that DiSCo unsuccessful trials always corresponded to timeouts, i.e., not completing the trajectory within $20$~s.
DiSCo produced faster trajectories towards the blocks, reducing the average time to complete a trial to $14.9$~s over No Copilot ($17.8$~s) and the State-Based Copilot ($19.97$~s).

\begin{table*}
\vspace{-1em}
    \caption{Comparison of user perceived experience on Block Choice and Drawers tasks. Arrows indicate the desirable direction: $\downarrow$ means lower is better, and $\uparrow$ means higher is better.}
    
    \begin{threeparttable}
    
    \begin{tabular}{c|c|ccccc}
    \toprule
    Task & Policy & Mental Demand~($\downarrow$) & Physical Demand~($\downarrow$) & Effort~($\downarrow$)& Performance~($\uparrow$)& Frustration~($\uparrow$)\\
    \midrule
    & No Copilot & 3.23 & 3.17 & 3.65 & 4.78 & 5.12 \\
    Block Choice & State-Based Copilot & 4.28 & 4.65 & 4.75 & 2.25 & 3.93 \\
    & DiSCo & \textbf{2.75} & \textbf{3.05} & \textbf{3.32} & \textbf{5.82} & \textbf{5.85} \\
    \midrule
    & No Copilot & 3.56 & \textbf{2.93} & 4.02 & 3.93 & \textbf{4.07} \\
    Drawers & State-Based Copilot & 4.53 & 4.32 & 4.90 & 1.70 & 2.22 \\
    & DiSCo & \textbf{3.28} & 3.00 & \textbf{3.41} & \textbf{3.95} & 4.05 \\
    
    \bottomrule
    \end{tabular}
    \end{threeparttable}
    
    \label{tab:survey}
\end{table*}

We summarily observed that DiSCo had the highest success rates, zero incorrect trials, faster trial times, and the best user-perceived ease, control, and autonomy (Appendix A, Table~\ref{tab:blocks}). 
Additionally, users found all-round improvements in perceived experience according to the extended survey questions adapted from the NASA Task Load
Index (Table~\ref{tab:survey}).
These results demonstrate that users both achieved higher performance and preferred to control DiSCo over the State-Based Copilot and No Copilot conditions.

\subsection{Robotic Arm Drawers}

We next evaluated the Drawers task, which is significantly harder than the Block Choice task because it is temporally extended, involving a sequence of actions that also require greater precision (e.g., opening the drawer).
In this more challenging task, we found the effects of DiSCo significantly pronounced in every category (Kruskal-Wallis Test, $p< 0.01$, Table~\ref{tab:drawers-stats}).

\begin{table*}[t!]
\vspace{-1em}
    \centering
    \caption{Comparison of performance on Drawers across all policy conditions.}

    \begin{threeparttable}

    \begin{tabular}{c|cccccccc}
        \toprule
        Policy &  Success & Correct Drawer & Correct Object & Time (s) & Collisions & Ease & Control & Effect \cr
        \midrule
        No Copilot &  0.28 & 0.89 & 0.67 & 93.03 & 9.11 & 2.66 & 2.82 & \textbf{3.16} \cr
        State-Based Copilot & 0.00 & 0.52 & 0.27 & 98.91 & 1.00 & 1.68 & 1.66 & 1.77 \cr
        DiSCo & \textbf{0.42} & \textbf{0.92} & \textbf{0.82} & \textbf{88.64} & \textbf{0.74} & \textbf{3.08} & \textbf{2.94} & 2.99 \cr
        \bottomrule
    \end{tabular}
    \end{threeparttable}

    \label{tab:drawer}
\end{table*}

\begin{table}[b!]
\centering
\caption{Kruskal–Wallis test results for \textbf{Drawers} dataset ($\alpha = 0.01$).
Statistics across $n=10$ participants.}
\label{tab:drawers-stats}
\begin{tabular}{lccc}
\toprule
Measure & $H$ & $p$ & $p < 0.01$ \\
\midrule
Correct Drawer & 15.620  & $4.1\times 10^{-4}$  & Yes \\
Correct Object & 15.062  & $5.4\times 10^{-4}$  & Yes \\
Success        & 16.870  & $2.2\times 10^{-4}$  & Yes \\
Time (s)       & 46.039  & $1.0\times 10^{-10}$ & Yes \\
Collisions     & 131.441 & $2.9\times 10^{-29}$ & Yes \\
Ease           & 67.279  & $2.5\times 10^{-15}$ & Yes \\
Control        & 74.085  & $8.2\times 10^{-17}$ & Yes \\
Effect         & 65.406  & $6.3\times 10^{-15}$ & Yes \\
Mental Demand         & 13.071  & $1.5\times 10^{-3}$  & Yes \\
Physical Demand    & 14.873  & $5.9\times 10^{-4}$  & Yes \\
Effort         & 14.846  & $6.0\times 10^{-4}$  & Yes \\
Performance    & 31.843  & $1.2\times 10^{-7}$  & Yes \\
Frustration    & 26.680  & $1.6\times 10^{-6}$  & Yes \\
\bottomrule
\end{tabular}
\end{table}

We observed that both the State-Based Copilot and DiSCo significantly reduced collisions in this task.
The No Copilot condition on average collided into the drawer, object, or desk an average of $9.11$ times per trial, but the State-Based Copilot only collided an average of $1.0$ times while DiSCo collided an average of $0.74$ times (Table~\ref{tab:drawer}, "Collisions"). 
However, the State-Based Copilot, across all 10 participants, achieved no successful trials where the entire sequence of (1) opening the correct drawer, (2) picking the correct object, (3) placing it in the correct drawer, was completed within the allotted time (Table~\ref{tab:drawer}, "Success"). 
In contrast, DiSCo achieved the highest success rate, achieving $42\%$ correct trials on this task.
DiSCo more accurately selected the correct drawer and object compared to No Copilot and State-Based Copilot.
Further, DiSCo achieved significantly faster trial times than both the No Copilot and State-Based Copilot conditions (Table~\ref{tab:drawer}, "Time").
Summarily, DiSCo was more accurate and faster, achieving the highest performance on the Drawers task in our user study.

Users perceived that DiSCo was easier to use and control (Table~\ref{tab:drawer}, "Ease" and "Control").
On average, users also reported that they felt their inputs more strongly affected the robotic arm's position in the No Copilot than DiSCo condition (Table~\ref{tab:drawer}, "Effect") suggesting they may have noticed that they were not fully in control of the robot's precise actions when using DiSCo.
Consistent with the Block Choice task, users found improvements in perceived ease of use in the extended NASA Task Load adapated survey questions (Table~\ref{tab:survey}).
Together, these results demonstrate that in a more challenging task involving a sequence of actions, DiSCo continued to significantly increase task performance.
This suggests that for more challenging temporally extended tasks, shared autonomy with DiSCo can improve user control and experience.

\section{Discussion}
\label{sec:discussion}

DiSCo combines shared autonomy with diffusion policy, where inpainting trajectories enables the visuomotor policy to generate actions more consistent with the user's actions.
DiSCo has interpretable hyperparameters: (1) $\gamma$ controls how much the user's action \textit{conforms} to the learned diffusion policy, (2) $\rho$ controls how \textit{consistent} the future policy actions are with past user inputs via inpainting, and (3) $\beta$ controls the real-time \textit{responsiveness} felt by the user.
Across multiple tasks, we observe that DiSCo outperforms baselines of No Copilot and a State-Based Copilot \cite{yoneda2023noise}.
Surprisingly, we observed that State-Based Copilots worsened performance relative to No Copilot in both Block Choice and Drawers.
This may be due to policy corruption, leading the copilot to infer the wrong user intent because it only processes the user action at time $t$.
In contrast, DiSCo considers $O$ frames of past user actions.
This longer horizon of user actions may help mitigate temporally brief errors in the user policy.

There is an important difference in state visitation between autonomous and shared autonomy robots. 
In shared autonomy, suboptimal user inputs may lead the arm to visit states unseen in the training demonstrations.
Consequently, we observed that diffusion policy models trained using fixed start or target positions did not perform well for shared autonomy.
Anecdotally, diffusion policy produced suboptimal actions in out-of-distribution states, leading to poor actions.
To address this, we randomized the starting or target positions in our robotic arm demonstrations, providing greater state variance during training. 
Empirically, we observed this conferred greater robustness for our tasks, and all results in this paper use models trained in this way.
Future work should improve the robustness and generalization of shared autonomy models to out-of-distribution shifts. 

Shared autonomy through DiSCo involves the policy updating the user action at every frame to help share control.
One benefit of having the user and policy intervene at every frame is the possibility of rapid motor feedback and correction, a hallmark of human motor proficiency and dexterity.
This stands in contrast to other assistive approaches, such as providing language commands to Vision Language Action (VLA) models or related models \cite{Duan2024-hl, Kim2024-xw, Zhi2024-us}. 
Speech commands can provide great assistive benefit for tasks that are more autonomous and do not require a rapid human-robot feedback loop, such as having a robot fold laundry.
Tasks requiring greater supervision and a rapid human-robot feedback loop may benefit more from shared autonomy that consider user input every time frame. 
In assistive applications that require responsiveness and short loop time, a further consideration is that VLAs may be more impractical due to their large model size and significant computational cost.

Our tasks were motivated by the BCI literature, where cursor control (like the vehicle in MetaDrive) and robotic arm control can meaningfully enable paralyzed participants to perform tasks. 
Future work may extend DiSCo for use with low-bandwidth inputs, including extending 2D inputs to higher degree-of-freedom control.

\textbf{Limitations:}
There are important limitations to address in future work.
First, due to the inference delay of diffusion policy, users could notice a small delay in the robotic arm being responsive to its inputs.
This study addressed this by blending user inputs, but we do not suggest this is optimal.
To minimize this lag, future work should evaluate methods to accelerate inference while not sacrificing quality, thereby reducing the overall loop time of the shared autonomy system.
Second, our study's observation horizon was $O=6$, meaning diffusion policy was conditioned on the last $600$~ms of observations.
However, long-horizon tasks may require memory over greater amounts of time.
We anticipate that future work may need to address the limitation in observation horizon, which will likely involve more than simply increasing $O$. 
Third, this work inherits limitations of imitation learning, including some brittleness to out-of-distribution states.
For example, we randomized initial and target states in training demonstrations to aid generalization to out-of-distribution states, an approach that may not be easily scalable to more complex tasks.
Future work may investigate other methods of learning a more robust autonomous policy, such as via RL agents (as in our MetaDrive experiments).

\textbf{Conclusion \& outlook:}
We present DiSCo, a method of shared autonomy with diffusion policy, that helps users increase user performance in robotic arm tasks under corrupted inputs.
In future work, we believe this method could be applied in systems where user control is challenging, including brain-computer interfaces \cite{Hochberg2012-ys, Collinger2013-aa, Wodlinger2015-ls, Lee2025-ud}, where the goal is to restore control to paralyzed people.
Future work may also extend this work to settings where user control is underactuated, and shared autonomy may help to process user inputs to carry out higher-dimensional tasks.

\begin{acks}
This work was supported by NIH DP2NS122037, NIH R01NS121097, NIH DP1HD121548, and the UCLA-Amazon Science Hub. J.C.K. is a co-founder of Luke Health, is on its Board of Directors, and has a financial interest in it. He is also the inventor of Intellectual Property owned by the UC Regents and licensed to Luke Health.
\end{acks}
\balance

\clearpage
\appendix
\section*{Appendix}
\label{sec:appendix}

\subsection*{A. Questionnaires Used in User Studies}

To quantify ease, control, and autonomy, we asked participants the following three questions after every trial.
\begin{itemize}
    \item \textit{Ease:} "How easy was the task?"
    \item \textit{Control:} "How in control did you feel when performing the task?"
    \item \textit{Effect:} "How much do you believe your inputs affected the trajectory of the robotic arm?"
\end{itemize}

Participants gave their response on a scale of $1$ to $5$, where $1$ was the most negative response and $5$ was the most positive response.

In the second phase of the study (with 3 participants for Block Choice task and 5 participants for Drawers task), we also administered five additional items adapted from the NASA Task Load Index to capture perceived mental demand, physical demand, effort, performance, and frustration, each rated on a 7-point scale ($1$ = very low, $7$ = very high except for frustration which is reversed).

\begin{itemize}
    \item \textit{Mental Demand:} "How mentally demanding was the task?"
    \item \textit{Physical Demand:} "How physically demanding was the task?"
    \item \textit{Effort:} "How hard did you have to work (mentally and physically) to accomplish your level of performance?"
    \item \textit{Performance:} "How successful were you in accomplishing what you were asked to do?"
    \item \textit{Frustration:} "How insecure, discouraged, irritated, stressed, or annoyed were you during the task?"
\end{itemize}

\subsection*{B. User Studies Participant Demographic Information}

\begin{table}[b!]
\centering
\caption{Participant demographics. \textbf{Exp.} denotes self-reported robotics experience:
(A) Master: collected data and trained successful robot policies;
(B) Intermediate: controlled robots for tasks (e.g., data collection or user studies);
(C) Novice: no experience controlling robotic arms.}

\begin{threeparttable}

\begin{tabular}{lccccl}
\toprule
\textbf{Task} & \textbf{ID} & \textbf{Age} & \textbf{Sex} & \textbf{Exp.} & \textbf{Ethnicity} \\
\midrule
Block         & 1  & 24 & M & C & Asian \\
Block         & 2  & 20 & M & C & Asian \\
Block         & 3  & 20 & F & C & Asian \\
Block         & 4  & 23 & F & C & Asian \\
Drawer        & 5  & 19 & M & C & Asian \\
Drawer        & 6  & 20 & F & C & Asian \\
Block (blend) & 7  & 22 & F & A & Asian \\
Drawer        & 8  & 20 & F & C & Asian \\
Drawer        & 9  & 24 & M & B & Asian \\
Block (blend) & 10 & 20 & M & A & Asian, White \\
Block (blend) & 11 & 27 & M & C & Other \\
Drawer        & 12 & 20 & M & C & Asian \\
Drawer        & 13 & 21 & M & B & Asian \\
Drawer        & 14 & 21 & F & A & White \\
Drawer        & 15 & 24 & M & B & Asian \\
Drawer        & 16 & 21 & F & B & Asian \\
Drawer        & 17 & 23 & M & C & Asian \\
Drawer        & 18 & 21 & F & B & Asian \\
Drawer        & 19 & 23 & M & B & Asian \\
Drawer        & 20 & 25 & M & B & White \\
Drawer        & 21 & 24 & F & B & Asian \\
\bottomrule
\end{tabular}
\end{threeparttable}

\label{tab:participant_demographics}
\end{table}

\clearpage
\bibliographystyle{ACM-Reference-Format}
\bibliography{refs}
\balance

\end{document}